\documentclass[twocolumn]{revtex4}
\usepackage{graphicx}
\newcommand{\beq}{\begin{equation}}
\newcommand{\enq}{\end{equation}}
\newcommand{\bea}{\begin{eqnarray}}
\newcommand{\ena}{\end{eqnarray}}
\newcommand{\ad}{a^{\dag}}
\begin{document}

\title{Calculation of collective modes for the Bose-Hubbard model
with confinement}
\author{Emil Lundh}
\affiliation{Department of Physics, Royal Institute of Technology,
Albanova, SE-106 91 Stockholm, Sweden}
\begin{abstract}
The collective excitations in the Bose-Hubbard model in a trap are
studied by means of numerical diagonalization in one dimension. The
strength function is calculated for monopole and dipole
perturbations, and moments of the strength function are utilized
in order to obtain information about the collective behavior under
external forces. In the superfluid regime, the spectrum is found
to be exhausted by one single frequency, while in systems that
contain a Mott insulating plateau several frequencies are
excited. An explanation of recent experimental findings in terms 
of a Mott plateau is suggested.
\end{abstract}
\maketitle
Recent years have seen a rapid development in the study of atoms
confined in optical lattices. One of the virtues of such systems
is that they provide an almost perfect realization of the bosonic
Hubbard
model. Corrections such as nearest-neighbor interactions 
are under control and can be made very small, and
furthermore, the parameters of the model are widely tunable
experimentally. In the case of spinless bosonic atoms the
Bose-Hubbard model is known to support a Mott transition, and it
has also been demonstrated experimentally by Greiner {\it et al.}
\cite{greiner2001a}
The addition of spatial confinement adds complexity to the
well-known Bose-Hubbard model. In these inhomogeneous systems 
superfluid and Mott insulating phases may coexist in
spatially separated regions \cite{jaksch1998a}. However, a good
experimental diagnostics is still lacking for the detection of this
interesting type of state.

This paper wishes to forward the idea of using oscillatory modes
in order to discriminate between different types of many-body state
for these systems. Exciting oscillatory modes is an experimentally
well-developed technique for studying the many-body physics of
trapped condensates in the absence of optical lattices and has
recently been applied also to the case with an optical lattice
present \cite{cataliotti,stoeferle2004a,schori2004a}.
Physical limits and basic dependences of the many-body eigenstates
and energies were studied in an earlier paper \cite{lundh2004a}
(cf.\ \cite{pupillo2003a}).
Here we wish to relate the eigenvalue spectrum to the collective
oscillations of the system and pose the physical question, what
happens when the system is perturbed? 
Just as in Ref.\ \cite{lundh2004a}, a five-particle system
shall be studied using exact diagonalization, but the methods and
concepts introduced here can be applied in numerical calculations
using methods such as Monte Carlo, density-matrix renormalization
group or mean-field calculations.

The system under consideration is a gas of bosonic atoms, confined
in a harmonic trap superposed with an optical lattice formed by
standing laser waves. By making the trap sufficiently tight in two
directions, the system is effectively made one-dimensional, and
when the optical-lattice potential is sufficiently strong the
system is well described by the Bose-Hubbard Hamiltonian
\beq\label{hamiltonian}
H = \frac12 U\sum_{r}\ad_r\ad_r a_r a_r -
\sum_{<rr'>}\ad_r a_{r'}
+\frac12 \omega_t^2\sum_{r}r^2 \ad_r a_r.
\enq
The units are here chosen so that the tunneling strength is unity.
$U$ is the on-site interaction strength, $\omega_t$ is the trap
frequency, and $r$ is a dimensionless site index; the operators
$a_r$,$\ad_r$ destroy and create a particle at the site $r$,
respectively, and obey Bose commutation relations. The sum
subscripted $<\!rr'\!>$ runs over all pairs of nearest-neighbor sites
as usual.

We wish to know how the system responds to a perturbation of the
type
\beq\label{perturbation}
F = \sum_r v(r) \ad_r a_r,
\enq
where $v(r)$ is a perturbing potential applied to the bosons.
The strength function associated with the operator $F$ is
\cite{textbook}
\beq
S(\omega) = \sum_{k \neq 0} |\langle k |F|0\rangle|^2
\delta(\omega-E_k),
\enq
where $k$ labels the many-body eigenstates of the system and
$E_k$ are the eigenenergies of the
many-body system (in units where $\hbar=1$, and defining the 
ground-state energy as the zero level).

In the present exact diagonalization study, the strength function
can be obtained exactly. However, considering the outlook of
using numerical methods where larger systems can be treated, but
the whole eigenvalue spectrum cannot easily be obtained,
approximate methods must be considered. Reference
\cite{lipparini1989a} describes a moment method (cf.\
\cite{bruun2001a,bohigas1979a}) that is well suited for describing
the excitation spectrum for confined bosons in optical lattices.
The moments of the strength function,
\beq
m_p = \int d\omega S(\omega) \omega^p,
\enq
can, namely, be written as ground-state expectation values of
commutators involving the Hamiltonian and the operator $F$. For
the three moments that shall be used here, $m_0$, $m_1$ and $m_2$,
one obtains
\bea
m_0 = \langle 0 | F^2|0\rangle - \langle 0 | F | 0 \rangle ^2,
\nonumber\\
m_1 = \frac12 \langle 0 |[F,[H,F]]|0\rangle,
\nonumber\\
m_2 = \langle 0 |[F,H][H,F]|0\rangle.
\ena
The moment $m_p$ has the dimension of frequency raised to the $p$th
power, times the dimension of $F^2$, and therefore the quotients
between different moments
define frequencies: from the ones listed above we define
$\omega_{10}=m_1/m_0$, $\omega_{20}=(m_2/m_0)^{1/2}$, and
$\omega_{21}=m_2/m_1$. In the case where only one eigenstate is
excited by $F$, these moment-related frequencies coincide and are equal to the
frequency of that state, but more generally, they
take on values between the frequencies of those
modes that are appreciably excited by $F$, and the spread between 
them is related to the spread of the strength function.
Observe that very little information about
the individual excitation frequencies can be obtained from the
moment calculation, except for the the fact that as $p\to-\infty$,
the frequencies $\omega_{p,p-1}$ approach the lowest-lying eigenstate 
with a symmetry consistent with the perturbation $F$ \cite{lipparini1989a}.
The three frequencies are, however, average excitation frequencies, in 
a sense that is made precise by considering their definition:
\beq
\omega_{10} = \frac{\sum_{k\neq 0}|\langle k |F|0\rangle|^2 E_k}{
\sum_{k\neq 0}|\langle k |F|0\rangle|^2},
\enq
and correspondingly for the other two frequencies.
Furthermore, by writing $m_0$ as
\beq
m_0=\sum_{k\neq 0}|\langle k |F|0\rangle|^2,
\enq
it is clear that the zeroth moment is related to the probability
for any excitation at all to happen, and it therefore gives
information about the amplitude of the oscillations excited by the
perturbation.

\begin{figure}
\includegraphics[width=\columnwidth]{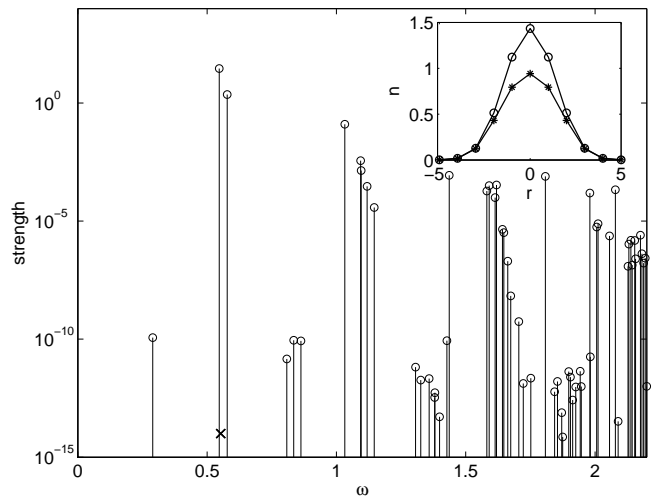}
\caption[]{Visualization of the strength function for a monopole
excitation of a weakly interacting system, with $\omega_t=0.3$,
$U=0.1$. The height of the stems represent the strength $S_{k0}$ 
for each excited state $|k\rangle$, and
their horizontal position indicates the frequency relative to the
ground state of the excited state in question, $\omega=E_k$.
The crosses mark the magnitude of three
moment-related frequencies, in increasing order, $\omega_{10}$,
$\omega_{20}$ and $\omega_{21}$. (However, the positions of the 
three crosses coincide completely in this figure.)
The inset shows the density profile as circles and the quantum 
fluctuations in the local density as asterisks.
All quantities are plotted in the dimensionless units employed
throughout the paper.
\label{fig:strengthf_weak}}
\end{figure}

Let us now explore in the setting of an exactly diagonalizable
system the relation between the strength function and the moment-related
frequencies $\omega_{pq}$. The details of the diagonalization scheme 
were explained in Ref.\ \cite{lundh2004a}. The system consists of $N=5$
particles in one dimension subject to the Hamiltonian, 
Eq.\ (\ref{hamiltonian}).
The perturbing potential is assumed to be on the form of Eq.\
(\ref{perturbation}) with the two choices of potential function
\bea
v(r) = r & \mbox{, dipole perturbation (tilting)},
\nonumber\\
v(r) = r^2 & \mbox{, monopole perturbation (compression).}
\ena
The exact strength function can be found by calculating the
strength $S_{k0}\equiv |\langle k | F | 0\rangle|^2$ for each excited state
$k$. The strength function for the finite system is a sum of
delta functions, and can of course not be plotted, but it is
visualized in Fig.\ \ref{fig:strengthf_weak} by plotting a stem of
height $S_{k0}$ at the position $E_k$ of each excited state.
For Fig.\ \ref{fig:strengthf_weak} the parameters are chosen as 
$\omega_t=0.3$, $U=0.1$, describing a weakly interacting system in a 
shallow trap,
and the perturbing potential $v$ is chosen as the monopole
perturbation. It is seen that only two almost-degenerate modes are 
excited with appreciable strength. These are the two lowest-lying  
monopole modes, $|2\rangle$ and $|3\rangle$, at $E_2\approx 0.5524$
and $E_3\approx 0.5776$; the excitation of the latter is weaker by 
approximately a factor 
10. The strength for the third-strongest excited state is weaker again 
by another factor 20. The strengths for the eigenstates that have dipole 
symmetry are suppressed by about ten orders of magnitude, which 
corresponds to the numerical accuracy of the algorithm.
This system is an almost pure Bose-Einstein
condensate, its density resembling a discretized Gaussian, as seen in 
the inset of Fig.\ \ref{fig:strengthf_weak}. 
Each of the system's collective modes is almost entirely exhausted by 
one many-body eigenstate or, as in the monopole case, an 
almost-degenerate pair of them.

Included in Fig.\ \ref{fig:strengthf_weak}, as crosses,
are the three moment-related frequencies $\omega_{pq}$, which can be
calculated using only ground-state properties.
It is seen that all three
frequencies lie close together -- the symbols coincide almost 
completely so that only one is visible in the plot -- and 
lie close to the main excitation frequency $E_2$, 
as expected. The magnitude of the zeroth
moment is $m_0=30.6515$; this number makes more
sense when compared to values of $m_0$ for other parameter
choices. It can be roughly interpreted to mean that application of
the potential $v(r)$, which means quadrupling the trap potential in 
this case,
yields a population transfer rate of more than order unity to excited 
states and the oscillations should be clearly visible.

\begin{figure}
\includegraphics[width=\columnwidth]{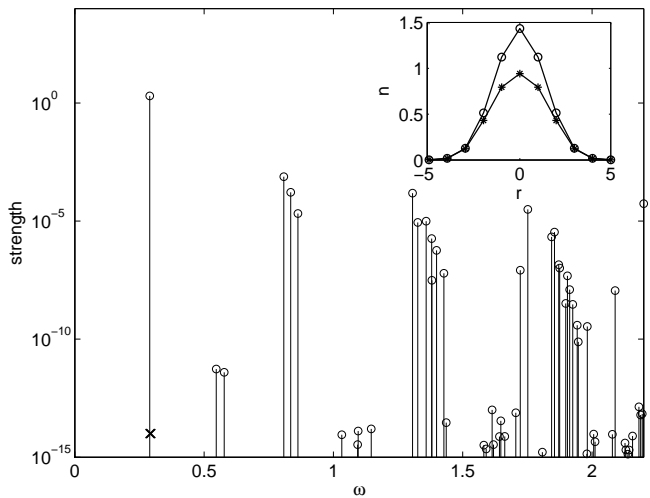}
\caption[]{Same as Fig.\ \ref{fig:strengthf_weak}, but here
the perturbation has dipole symmetry.
\label{fig:strengthf_dipole}}
\end{figure}

Figure \ref{fig:strengthf_dipole} displays the results for the dipole
excitation using the same parameters $\omega_t=0.3$, $U=0.1$.
The picture is similar, but even simpler than in the monopole case: 
only one mode is appreciably excited, which in physical terms means 
that a perturbation with dipole symmetry
results in excitation of the lowest eigenstate of the many-body
system, and thus oscillation at one well-defined frequency. The
zeroth moment is $m_0=1.9928$, again of order unity.

\begin{figure}
\includegraphics[width=\columnwidth]{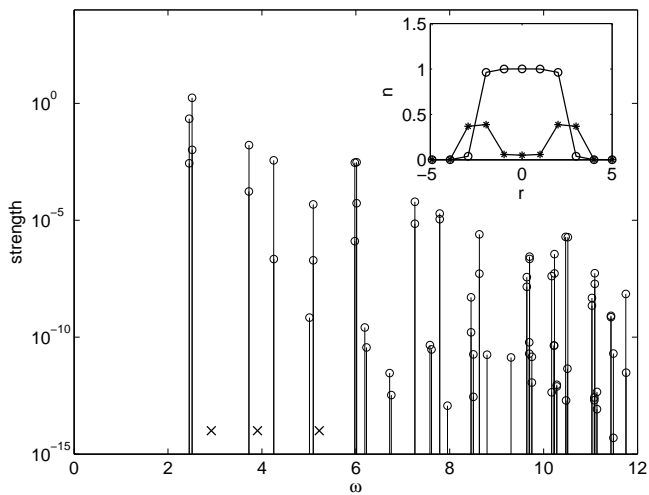}
\caption[]{Same as Fig.\ \ref{fig:strengthf_weak}, but here
the trap strength is $\omega_t=1$ and the coupling is intermediate, $U=20$.
In the inset, the asterisks represent the quantum fluctuations
in the density magnified by a factor 10.
\label{fig:strengthf_intermediate}}
\end{figure}

Another example is shown in Fig.\ \ref{fig:strengthf_intermediate}.
The parameter choice $\omega_t=1$, $U=20$ corresponds to a mixed state
with a rudimentary Mott insulating plateau in the center. At the
present there is very little understanding of the oscillatory
modes in this type of regime and little hope for an analytical
approach. The results for the strength function is displayed in
the main panel of Fig.\ \ref{fig:strengthf_intermediate}. Clearly, the
strength is now spread out over many eigenstates, although the
largest strength is about a factor 10 larger than the
next-largest. In experiment one can expect to observe a monopole
oscillation with several frequencies, if damping is weak enough
and the precision is good enough to resolve the different
frequencies.
The results of the moment calculation are likewise included in
Fig.\ \ref{fig:strengthf_intermediate}. The three moment-related 
frequencies now differ appreciably, as expected from the spread of the
excitation strengths. As noted above, the moment calculation does not 
give direct information about the individual frequencies that are 
excited but the $\omega_{pq}$ lie between the frequencies of the 
most strongly excited modes.
The zeroth moment for the Mott-plateau system is found to be
$m_0=1.9596$. This indicates that again, excitation should result in
visible oscillations.

\begin{figure}
\includegraphics[width=\columnwidth]{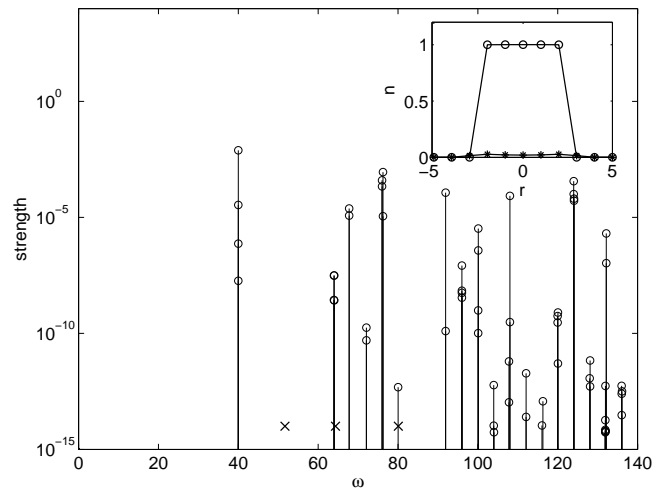}
\caption[]{Same as Fig.\ \ref{fig:strengthf_weak}, but here
the trap and the coupling are stronger, $\omega_t=4$ and $U=100$.
In the inset, the asterisks represent the quantum fluctuations
in the density magnified by a factor 100.
\label{fig:strengthf_strong}}
\end{figure}

The final example is in the limit of strong coupling
in a tight trap, $U=100$ and $\omega_t=4$. This system is Mott
insulating \cite{lundh2004a}.
Figure \ref{fig:strengthf_strong}
depicts the strength function and the moment-related frequencies
for this system.
Just as in the Mott-plateau case, the
strength is distributed over several modes, which shows up as a
difference between the moment-related frequencies. Observe,
however, the magnitude of the strength function: correspondingly,
the zeroth moment is now very small, $m_0=0.0094$. The reason is
that the ground state for this system is a pure Fock state in
configuration space, and therefore an eigenstate of the
monopole operator and in fact of any operator of the form of Eq.\
(\ref{perturbation}), with very small corrections. This is of
course a manifestation of the insulating nature of this state. 

\begin{figure}
\includegraphics[width=\columnwidth]{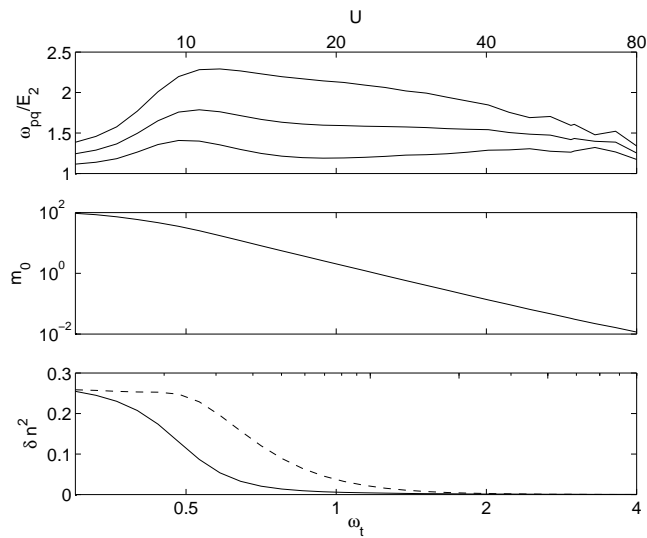}
\caption[]{Properties of the system as 
functions of the coupling $U$ and the trap strength $\omega_t$. 
The latter has been swept from $\omega_t=0.3$ to $\omega_t=4$, and
and $U$ has varied as $U=20\omega_t$ through the sweep. 
The topmost panel shows 
the three moment-related
frequencies (from top to bottom $\omega_{10}$,
$\omega_{20}$ and $\omega_{21}$) in units of the
lowest monopole frequency $E_2$.
The middle panel shows
the magnitude of the zeroth moment $m_0$.
The bottom panel plots the quantum fluctuations in local density,
$\delta n^2 = \langle n^2\rangle - \langle n \rangle^2$, at the
point $r=0$ (solid line) and $r=2$ (dashed line). The units are
the dimensionless units employed throughout the paper.
\label{fig:momentsweep}}
\end{figure}

The final figure is an overview picture. Figure
\ref{fig:momentsweep} displays the three
moment-related frequencies in units of the lowest monopole 
frequency $E_2$; the zeroth moment $m_0$; 
and the quantum fluctuations in the
local density at two points, all as functions of the coupling $U$
and the trap frequency $\omega_t$.
The data has been recorded along an oblique path through phase 
space along the line $U=20\omega_t$, where $\omega_t$ has varied 
from the value 0.3 up to 4; this way, all the phases discussed
above could be sampled.
For small $U$ and $\omega_t$,  the system is superfluid and
the density fluctuations are large. The
moment-related frequencies stay close to each other and to the
eigenfrequency $E_2$. 
The decrease of the density
fluctuations in the centermost well 
signals the transition to a 
Mott-plateau state, and at the same time the frequencies begin to
differ, but the zeroth moment is still above unity. 
The drop in $m_0$
comes at the same point as the drop in the fluctuations in the
second off-center site;
this marks the transition to the Mott insulating state. In this
finite system, the Mott transition is gradual.

Reference \cite{schori2004a} reports on an experiment that 
measures the response to excitation of
trapped bosons in a three-dimensional optical lattice. 
It is found that 
for a shallow lattice, which corresponds to a relatively small 
$U$ compared with the tunneling strength, one narrow frequency range 
is excited, just as expected for a trapped condensate. However, 
for larger $U$, a broad range of frequencies is excited with 
appreciable strength. This result, which is not easily interpreted 
in the framework of Bogoliubov theory, is in fact very 
similar to the present findings for a system with a Mott plateau. 
The drastic difference in size, dimensionality and method 
of excitation forbids
quantitative comparison, but the qualitative similarities are there: 
we have seen in this paper that 
the rise and broadening of the frequencies is 
characteristic of being in the vicinity of the Mott-plateau regime.

In conclusion, the strength function for the monopole and dipole
excitations of the Hubbard model in a trap have been calculated,
and with an outlook to treating larger systems, the moments of the
strength function have been employed in order to extract
information about the excitation strengths and frequencies. It is
seen in both the moment calculations and the full strength
function how the finite system undergoes a transition between
superfluid, Mott plateau, and fully Mott insulating states. It is
predicted that the experimental signature of a system containing a
Mott plateau is the excitation of several frequencies with
appreciable strength, while a strongly coupled system will not be
excited at all because of its Mott insulating nature. Based on 
the present findings, it is suggested that the broad and strong
resonance found in a recent experiment \cite{schori2004a} is due 
to a Mott insulating plateau in the system.

\begin{acknowledgments}
This project was financially supported by the G{\"o}ran Gustafsson
foundation. 
\end{acknowledgments}




\begin{thebibliography}{99}
\bibitem{greiner2001a} M.\ Greiner, O.\ Mandel, T.\ Esslinger,
  T.\ W.\ H{\"a}nsch, and I.\ Bloch, Nature {\bf 415}, 39 (2002).
\bibitem{jaksch1998a} D.\ Jaksch, C.\ Bruder, J.\ I.\ Cirac,
  C.\ W.\ Gardiner, and P.\ Zoller, Phys.\ Rev.\ Lett.\ {\bf 81},
  3108 (1998).
\bibitem{cataliotti} S.\ Burger {\it et al.}, Phys.\ Rev.\ Lett.\
  {\bf 86}, 4447 (2002);
  F.\ S.\ Cataliotti {\it et al.}, Science {\bf 293}, 843 (2001).
\bibitem{stoeferle2004a} T.\ St{\"o}ferle, H.\ Moritz, C.\ Schori,
  M.\ K{\"o}hl, and T.\ Esslinger, Phys.\ Rev.\ Lett.\ {\bf 92},
  130403 (2004).
\bibitem{schori2004a} C.\ Schori, T.\ St{\"o}ferle, H.\ Moritz,
  M.\ K{\"o}hl, and T.\ Esslinger, e-print cond-mat/0408449 (2004).
\bibitem{lundh2004a} E.\ Lundh, Phys.\ Rev.\ A, in press.
\bibitem{pupillo2003a} G.\ Pupillo, E.\ Tiesinga, and
  C.~J.\ Williams, Phys.\ Rev.\ A {\bf 68}, 063604 (2003).
\bibitem{textbook} A.\ Bohr and B.~R.\ Mottelson, 
  {\it Nuclear Structure} (World Scientific, Singapore, 1998), 
  Vol.~1.
\bibitem{lipparini1989a} E.\ Lipparini and S.\ Stringari, 
  Phys.\ Rep.\ {\bf 175}, 103 (1989).
\bibitem{bruun2001a} G.~M.\ Bruun, Phys.\ Rev.\ A {\bf 63}, 
  043408 (2001).
\bibitem{bohigas1979a} O.\ Bohigas, A.~M.\ Lane, and J.\ Martorell, 
  Phys.\ Rep.\ {\bf 51}, 267 (1979).
%
\end{thebibliography}
\end{document}